# Nonlinear effects in spin relaxation of cavity polaritons

D.D. Solnyshkov [1], I.A. Shelykh[2,3], M.M. Glazov[4], G. Malpuech[1], T. Amand [5], P. Renucci[5], X. Marie[5], and A.V Kavokin[3]

[1]*LASMEA, UMR 6602 CNRS, Université Blaise-Pascal, 24, av des Landais, 63177, Aubiere, France*

[2]*St.Petersburg State Polytechnical University, Polytechnicheskaya ul. 29, 195251, St. Petersburg, Russia*

[3]*Physics and Astronomy School, University of Southampton, Highfield, Southampton, So17 1BJ, United Kingdom*

[4] *A.F. Ioffe Physico-Technical Institute, 26 Polytechnicheskaya, St Petersburg, Russia.*

[5]*Laboratoire de Magnétisme Nanophysique et Optoélectronique, INSA, 135 Avenue de Rangueil, 31077 Toulouse Cedex, France*

PACS :71.36.+c, 71.35.Lk, 78.47.+p, 42.50.Md

**Abstract.** We present the general kinetic formalism for the description of spin and energy relaxation of the cavity polaritons in the framework of the Born-Markov approximation. All essential mechanisms of polaritons redistribution in reciprocal space together with final state bosonic stimulation are taken into account, from our point of view. The developed theory is applied to describe our experimental results on the polarization dynamics obtained in the polariton parametric amplifier geometry (pumping at so-called *magic angle*). Under circular pumping, we show that the spin relaxation time is strongly dependent on the detuning between exciton and cavity mode energies mainly because of the influence of the detuning on the coupling strength between the photon-like part of the exciton-polariton lower dispersion branch and the reservoir of uncoupled exciton states. In the negative detuning case we find a very long spin relaxation time of about 300 ps. In the case of excitation by a linearly polarized light, we have experimentally confirmed that the anisotropy of the polariton-polariton interaction is responsible for the build up of the cross-linear polarisation of the signal. In the spontaneous regime the polarisation degree of the signal is –8% but it can reach –65% in the stimulated regime. The long-living linear polarization observed at zero detuning indicates that the reservoir is formed by excitons localized at the anisotropic islands oriented along the crystallographic axes. Finally, under elliptical pumping, we have directly measured in the time domain and modelled the effect of self-induced Larmor precession, i.e. the rotation of the linear polarisation of a state about an effective magnetic field proportional to the projection of the total spin of exciton-polaritons in the cavity on its growth axis.

# 1. Introduction

In the last decade quantum microcavities in the strong coupling regime were quite intensively studied. The elementary excitations of such systems are exciton-polaritons [1] which possess a number of unusual physical properties inherited from both bare excitons and cavity photons. Similarly to excitons, cavity polaritons obey bosonic statistic if their 2D concentration lies below the saturation density $n_{sat} \sim a_B^{-2}$, where $a_B$ is the 2D exciton Bohr radius. Experimentally, the final state bosonic stimulation of polariton relaxation has been documented [2]. Due to the presence of the photonic component, exciton polaritons have an extremely small effective mass at k=0 (about $10^{-4}$ of the free electron mass in vacuum). Combined with the efficient energy relaxation this makes them ideal candidates for observation of the Bose condensation at high temperatures [3,4].

The in-plane dispersion of the cavity polaritons is substantially different from the dispersions of the bare excitons and cavity photons. It strongly depends on the energy difference between the bottoms of the photonic and excitonic bands in the cavity refered to as the detuning parameter $\delta$. The detuning is positive if the bottom of the photonic band lies above the excitonic band. In this case the dispersion curves of the upper and lower polaritons are almost parabolic and resemble those of the photons and bare excitons. On the other hand, if the bottom of the photonic band lies below the bottom of the excitonic band, the anticrossing of upper and lower polariton branches leads to the strong non parabolicity of the polariton dispersion. In the vicinity of resonance between bare exciton and photon energies, the polariton effective mass varies drastically between the masses of a cavity photon and of a bare exciton. This leads to formation of the bottleneck region in the lower polariton branch (LPB), which efficiently suppresses the polariton energy relaxation assisted by acoustic phonon scattering [5]. To overcome the bottleneck effect, one should either use doped samples where the free carriers are present [6], or work in the strong pumping regime where the polariton concentration is high enough and polariton-polariton scattering is efficient.

Under strong optical pumping, the stimulated polariton-polariton scattering affects optical properties of microcavities. Non-linear effects are extremely pronounced in the polariton parametric amplifier configuration [7,8], where the LPB is resonantly excited under the so-called magic angle $\theta_p$, so that that the pair of the pumped polaritons scatters into the signal (ground) and idler states. To make this process efficient, the energy and momentum should be conserved simultaneously, $2k_p = k_i$, $2E_{LP}(k_p) = E_{LP}(0) + E_{LP}(k_i)$. These conditions can be satisfied in microcavities in the strong coupling regime due the non-parabolicity of LPB. $k_p$ defines the so called *magic angle*. It is close to the inflexion point of the LPB at zero detuning. Due to the effect of bosonic stimulation the intensity of the parametric scattering can be sufficiently increased by shining the sample by a weak probe pulse incident normally to the structure interface [2].

In III-V zinc-blend semiconductor quantum wells (QWs), the lowest energy level of a heavy hole lies typically lower than any light-hole energy level, and thus the exciton ground state is formed by an electron and a heavy-hole. The total exciton spin in a QW has $\pm 1$ and $\pm 2$ projections on the structure axis. The states $|\pm2\rangle$ are not coupled to light and do not participate in polariton formation, while the states $|\pm1\rangle$ form an optically active polariton doublet. They can be created by $\sigma^+$ and $\sigma^-$ circularly polarized light, respectively. Linearly polarized light excites a linear combination of +1 and –1 exciton states, so that the average exciton spin projection to the structure axis is zero in this case while the exciton dipole moment has a non-zero in-plane component. As any two-level system, the optically active exciton doublet can be characterized by a pseudospin whose projections on the coordinate axes can be expressed via the elements of excitonic density matrix. The normal-to-plane pseudospin component characterizes the excitonic spin polarization, while its in-plane components describe the exciton dipole moment orientation.

The initial polarization of the polariton system inevitably decreases due to various spin relaxation processes. For quantum well excitons the most probable mechanism of spin relaxation is due to the long range electron-hole exchange interaction (Bir- Aronov- Pikus mechanism [9]), which splits the excitonic states with dipole moment perpendicular and parallel to their wavevector [10]. This splitting referred to as longitudinal-transverse (LT) splitting or TE-TM splitting in microcavities creates an effective in-plane magnetic field acting on the polariton's pseudospin and leading to spin relaxation within the optically active polariton doublet $\{|+1\rangle,|-1\rangle\}$ related to the motional narrowing effect [10]. The absolute value of this splitting is zero for the ground state (k=0) and normally increases with the wave number $k$ [11]. In quantum microcavities the LT splitting is enhanced due to the coupling with photonic modes, whose energies are different in TE and TM polarizations at non-zero in-plane wave-vectors.

Pseudospin dynamics of cavity polaritons is closely connected with their energy relaxation. The final state bosonic stimulation plays a crucial role in both processes. It was first demonstrated experimentally by studying the temporal dependence of the circular polarization degree $\rho_c$ of the photoluminescence from the ground state of the LPB in a II-VI microcavity [12] under non-resonant excitation. Below the stimulation threshold the degree of circular polarization of emitted light $\rho_c$ exhibited an exponential decay, while above the threshold pronounced oscillations of $\rho_c$ have been observed,. This was explained theoretically in [13] within a framework of the pseudospin formalism. It was argued that bosonic stimulation is crucial for the observed effect. Indeed, below the threshold, the spin system is in the collision dominated regime, i.e. relaxation of polaritons down to the ground state goes through a huge number of random passes each corresponding to the scattering act with an acoustic phonon. In each scattering event the direction of the effective magnetic field changes randomly, so that in average no oscillation of any of pseudospin projections can be observed. The situation is fully analogous to one for electrons undergoing spin relaxation in the effective spin-orbit field [14]. However, after passing the stimulation threshold the scattering is no more random. Due to the effect of bosonic stimulation it is "directed" from the pump region to the bottleneck, where polaritons are accumulated and have enough time to rotate their pseudospins around the direction of magnetic field produced by the LT splitting. The oscillations of circular polarization degree of the polarization at the ground polariton state reflect precession of the pseudospin in the bottleneck region.

The model presented in [13] neglected polariton-polariton interactions. This made it unsuitable for the description of the polarization dynamics of the polariton parametric amplifiers, which reveals some intriguing properties. Experimentally it has been studied in Ref.[15]. Quite surprisingly, this work demonstrated that for the linear polarized pump and circular probe the signal is also linearly polarized, but its polarization plane is rotated by about 45° with respect to the pump polarization. In [16] this was qualitatively explained as a result of the self-induced Larmor precession of the pseudospins of polaritons resonantly excited by the pumping pulse. The proposed scenario was as follows. The circular probe enhances the parametric scattering of the polaritons with spins parallel to those of polaritons created by the probe pulse. This introduces an imbalance between spin-up and spin-down polaritons in the pump state. Due to the anisotropy of the polariton- polariton interactions it leads to the $\sigma^+/\sigma^-$ splitting in the pump state [15], which results in the rotation of the in-plane polariton polarisation of the pump in a direction which depends on the sign of $\rho_c$ (the self-induced Larmor precession). This rotation results in oscillating linear polarization of the emission from the ground state.

As in [13], the kinetic model of [16] had a phenomenological character. Besides, it took into account only scattering of polaritons in triplet configuration (parallel spins), ignoring the possibility of singlet scattering since it is usually much less effective [17]. However, recently it was argued that it can play an important role in the parametric polariton scattering and can lead to the orthogonal linear polarizations of the signal and pump polaritons [18,19]. Another important effect which has recently been evidenced is the splitting of polariton eigen-states into a doublet polarized along the crystallographic axes [110] and [1-10] [19] which is most probably related with additional exchange splitting of the exciton states localized at one or another interface of the quantum well.

In this article, we present a microscopic formalism for description of the spin and energy relaxation of cavity polaritons in the nonlinear regime. It is based on the Markovian treatment of the Liouville- von Neumann equation for the density matrix and takes into account all above mentioned mechanisms of the redistribution of polaritons in the reciprocal space, including polariton-polariton scattering in both singlet and triplet configuration. This theory is applied for description of a full set of the experimental data on the polarization dynamics of polaritons in the parametric amplifier geometry for different values of the detuning. Unlike Ref. 15 we had no probe in our experiment and resolved temporally the signal emission detecting its polarization for different polarizations of pump. For circularly polarized excitation, we observed a pronounced detuning effect on the polariton spin-relaxation: the life time of circular polarization of the signal is much longer for negative than for positive detunings. The set of our data taken at elliptically polarized and linearly polarized excitation, has been recently published [18] but only interpreted qualitatively within a simplified theoretical model. Here we reproduce theoretically all the essential features of the experiment, particularly the temporal oscillations of the linear polarization degree for the elliptical pump and inversion of the linear polarization of the signal for the linear pump.

The paper is organised as follows. In Section 2 we present in detail our theoretical model. In Section 3 we present experimental results and compare them with the results of the theoretical calculation.

## 2. Theoretical model

In this Section we derive the kinetic equations describing both spin and energy relaxation of the exciton polaritons. We start from the well-known Hamiltonian for interacting excitons and photons written in the basis of the circular polarized components

$$H = \sum_{\mathbf{k},\sigma} \left( \varepsilon_{ex,\mathbf{k}} a_{\mathbf{k}\sigma}^+ a_{\mathbf{k}\sigma} + \varepsilon_{phot,\mathbf{k}} c_{\mathbf{k}\sigma}^+ c_{\mathbf{k}\sigma} + \varepsilon_{phon,\mathbf{k}} b_{\mathbf{k}\sigma}^+ b_{\mathbf{k}\sigma} \right) + \frac{V_{Rabi}}{2} \sum_{\mathbf{k},\sigma} \left( a_{\mathbf{k}\sigma}^+ c_{\mathbf{k}\sigma} + a_{\mathbf{k}\sigma} c_{\mathbf{k}\sigma}^+ \right) + \hbar \sum_{\mathbf{k},\sigma} \left( \Omega_{ex,\mathbf{k}} a_{\mathbf{k}\sigma}^+ a_{\mathbf{k}-\sigma} + \Omega_{ph,\mathbf{k}} c_{\mathbf{k}\sigma}^+ c_{\mathbf{k}-\sigma} \right) +$$

$$+ \sum_{\mathbf{k},\mathbf{q},\sigma} U^{ex}_{\mathbf{k},\mathbf{q}} a_{\mathbf{k}\sigma}^+ a_{\mathbf{k}-\mathbf{q},\sigma} \left( b_{\mathbf{q}} + b_{-\mathbf{q}}^+ \right) + \frac{1}{4} \sum_{\mathbf{k},\mathbf{k}',\mathbf{q},\sigma} \left[ V^{ex(1)}_{\mathbf{k},\mathbf{k}',q} a_{\mathbf{k},\sigma}^+ a_{\mathbf{k}',\sigma}^+ a_{\mathbf{k}-\mathbf{q},\sigma} a_{\mathbf{k}'+\mathbf{q},\sigma} + 2 V^{ex(2)}_{\mathbf{k},\mathbf{k}',q} a_{\mathbf{k},\sigma}^+ a_{\mathbf{k}',-\sigma}^+ a_{\mathbf{k}-\mathbf{q},\sigma} a_{\mathbf{k}'+\mathbf{q},-\sigma} \right] + H.c. \tag{1}$$

where the index $\sigma$ denotes the spin projection of the excitons and photons on the structure growth axis, annihilation operators $a_{\mathbf{k}\sigma}$ correspond to the excitons, operators $c_{\mathbf{k}\sigma}$ describe annihilation of photons and $b_{\mathbf{k}}$ are annihilation operators of the acoustic phonons. The first term describes free particles propagation, the second accounts for the coupling between excitons and cavity photons, the third term describes the LT splitting of the bright excitonic doublet and the cavity mode, the fourth term describes exciton-acoustic phonon scattering, the fifth term describes the exciton-exciton scattering. The coefficient $V^{ex(1)}_{\mathbf{k},\mathbf{k}',q}$ is an amplitude of scattering of two excitons having the same spin projection on the structure growth axis (triplet scattering), $V^{ex(2)}_{\mathbf{k},\mathbf{k}',q}$ corresponds to the scattering of the two excitons having opposite spin projections (singlet scattering). If the scattering is spin-isotropic, $V^{ex(1)}_{\mathbf{k},\mathbf{k}',q} = V^{ex(2)}_{\mathbf{k},\mathbf{k}',q}$, but usually it is not the case for 2D excitons. The exchange interaction dominates exciton-exciton interaction in quantum wells [17], and consequently $\left| V^{ex(1)}_{\mathbf{k},\mathbf{k}',q} \right| >> \left| V^{ex(2)}_{\mathbf{k},\mathbf{k}',q} \right|$. In (1) we have neglected the dark polariton states as they cannot be optically created and can be populated either due to the spin relaxation of the single electrons and holes, or due to the exchange process : $\left| +1 \right\rangle \left| -1 \right\rangle \Rightarrow \left| +2 \right\rangle \left| -2 \right\rangle$ [20]. This approximation is certainly valid for detuning negative or zero ($\delta \le 0$) because of the lack of energy conservation in scattering from bright polaritons to dark excitons [18] (at $k$=0 the latter have higher energies, typically $\sim V_{Rabi} / 4$ above the ground state of the LPB).

In the strong coupling regime, it is convenient to use the polariton basis, introducing the annihilation operators of the upper and lower polaritons

$$p_{U,\mathbf{k}\sigma} = C_{U,\mathbf{k}} c_{\mathbf{k}\sigma} + X_{U,\mathbf{k}} a_{\mathbf{k}\sigma}$$
$$p_{L,\mathbf{k}\sigma} = C_{L,\mathbf{k}} c_{\mathbf{k}\sigma} + X_{L,\mathbf{k}} a_{\mathbf{k}\sigma} \tag{2}$$

where $C_{U/L,\mathbf{k}}, X_{U/L,\mathbf{k}}$ are Hopfield coefficients describing the photonic and excitonic fraction of the polariton quantum state, respectively, which depend on the cavity Rabi splitting and detuning parameter. Retaining the operators corresponding to the LPB only, since we want to describe the resonant excitation of LPB and neglecting the saturation effects, the Hamiltonian can be rewritten as [21] :

$$H = \sum_{\mathbf{k},\sigma} E(\mathbf{k}) p_{\mathbf{k}\sigma}^+ p_{\mathbf{k}\sigma} + \hbar \sum_{\mathbf{k},\sigma} \omega_{\mathbf{k}} b_{\mathbf{k}}^+ b_{\mathbf{k}} + \hbar \sum_{\mathbf{k},\sigma} \Omega_{LT,\mathbf{k}} p_{\mathbf{k}\sigma}^+ p_{\mathbf{k}-\sigma} +$$

$$+ \sum_{\mathbf{k},\mathbf{q},\sigma} U_{\mathbf{k},\mathbf{q}} p_{\mathbf{k},\sigma}^+ p_{\mathbf{k}-\mathbf{q},\sigma} \left( b_{\mathbf{q}} + b_{-\mathbf{q}}^+ \right) + \frac{1}{4} \sum_{\mathbf{k},\mathbf{q},\sigma} \left[ V^{(1)}_{\mathbf{k},\mathbf{k}',q} p_{\mathbf{k},\sigma}^+ p_{\mathbf{k}',\sigma}^+ p_{\mathbf{k}-\mathbf{q},\sigma} p_{\mathbf{k}'+\mathbf{q},\sigma} + 2 V^{(2)}_{\mathbf{k},\mathbf{k}',q} p_{\mathbf{k},\sigma}^+ p_{\mathbf{k}',-\sigma}^+ p_{\mathbf{k}-\mathbf{q},\sigma} p_{\mathbf{k}'+\mathbf{q},-\sigma} \right] + H.c. \tag{3}$$

where $E\left(\mathbf{k}\right) = \left(\varepsilon_{ex,\mathbf{k}} + \varepsilon_{ph,\mathbf{k}} - \sqrt{\left(\varepsilon_{ex,\mathbf{k}} - \varepsilon_{ph,\mathbf{k}}\right)^2 + V_{Rabi}^2}\right)\Big/2$ is the dispersion relation for LPB, neglecting the

TE-TM splitting, $\Omega_{LT,\mathbf{k}} = \Omega_{ex,\mathbf{k}}\left|X_{L,\mathbf{k}}\right|^2 + \Omega_{phot,\mathbf{k}}\left|C_{L,\mathbf{k}}\right|^2$, $U_{\mathbf{k},\mathbf{q}} = U^{ex}_{\mathbf{k},\mathbf{q}} X^*_{\mathbf{k}} X_{\mathbf{k-q}}$,

$V^{(i)}_{\mathbf{k},\mathbf{k}',\mathbf{q}} = V^{ex(i)}_{\mathbf{k},\mathbf{k}',\mathbf{q}} X^*_{\mathbf{k+q}} X^*_{\mathbf{k}'-\mathbf{q}} X_{\mathbf{k}} X_{\mathbf{k}'}$, $i$=1,2.

Now let us pass to the interaction representation and rewrite the Hamiltonian (3) as a sum of the "shift" and "scattering" terms [22]

$$H = H_{shift}(t) + H_{scatt}(t), \tag{4}$$

Here the "shift" term describes interaction of exciton-polaritons without wavevector transfer ($\mathbf{q}$=0) but having possibly different spins:

$$
\begin{aligned}
H_{shift} &= \sum_{\mathbf{k},\sigma=\uparrow,\downarrow} \left( \hbar\Omega_{LT,\mathbf{k}} p^+_{\sigma,\mathbf{k}} p_{-\sigma,\mathbf{k}} + V^{(1)}_{\mathbf{k},\mathbf{k},0}\left(p^+_{\sigma,\mathbf{k}} p_{\sigma,\mathbf{k}}\right)^2 + V^{(2)}_{\mathbf{k},\mathbf{k},0} p^+_{\sigma,\mathbf{k}} p^+_{-\sigma,\mathbf{k}} p_{\sigma,\mathbf{k}} p_{-\sigma,\mathbf{k}} \right) + \\
&+ \sum_{\mathbf{k}\neq\mathbf{k}',\sigma=\uparrow,\downarrow}\left( V^{(1)}_{\mathbf{k},\mathbf{k}',0} p^+_{\sigma,\mathbf{k}} p^+_{\sigma,\mathbf{k}'} p_{\sigma,\mathbf{k}} p_{\sigma,\mathbf{k}'} + V^{(2)}_{\mathbf{k},\mathbf{k}',0} p^+_{\sigma,\mathbf{k}} p^+_{-\sigma,\mathbf{k}'} p_{\sigma,\mathbf{k}} p_{-\sigma,\mathbf{k}'} \right)
\end{aligned}
\tag{5}
$$

The "scattering" term describes scattering between states with wave vector transfer ($\mathbf{q}\neq0$) :

$$
H_{scatt} = \frac{1}{4}\sum_{\mathbf{k},\mathbf{k}',\mathbf{q}\neq0,\sigma=\uparrow,\downarrow} e^{\frac{it}{\hbar}(E(\mathbf{k})+E(\mathbf{k}')-E(\mathbf{k+q})-E(\mathbf{k}'-\mathbf{q}))} \{ V^{(1)}_{\mathbf{k},\mathbf{k}',\mathbf{q}} p^+_{\sigma,\mathbf{k+q}} p^+_{\sigma,\mathbf{k}'-\mathbf{q}} p_{\sigma,\mathbf{k}} p_{\sigma,\mathbf{k}'} +
$$

$$
+2V^{(2)}_{\mathbf{k},\mathbf{k}',\mathbf{q}} p^+_{\sigma,\mathbf{k+q}} p^+_{-\sigma,\mathbf{k}'-\mathbf{q}} p_{\sigma,\mathbf{k}} p_{-\sigma,\mathbf{k}'} \} + \frac{1}{2}\sum_{\mathbf{k},\mathbf{q}\neq0,\sigma=\uparrow,\downarrow} U_{\mathbf{k},\mathbf{q}} \left( b_{\mathbf{q}} + b^+_{-\mathbf{q}} \right) p^+_{\sigma,\mathbf{k+q}} p_{\sigma,\mathbf{k}} \left( e^{\frac{it}{\hbar}(E(\mathbf{k})+\varepsilon_{phon,\mathbf{q}}-E(\mathbf{k+q}))} + e^{\frac{it}{\hbar}(E(\mathbf{k})-\varepsilon_{phon,-\mathbf{q}}-E(\mathbf{k+q}))} \right) + H.c.
$$

$$\tag{6}$$

With use of the Hamiltonian (4) we write the Liouville equation for the total density matrix of the system $\rho$ :

$$i\hbar\frac{d\rho}{dt} = \left[H(t),\rho\right] = \left[H_{shift}(t) + H_{scatt}(t),\rho\right]. \tag{7}$$

Eq. (7) can be treated within the Born-Markov approximation familiar in the quantum optics [23]. The Markov approximation means physically that the system is assumed to have no phase memory. It is, in general, not true for the coherent processes described by the Hamiltonian $H_{shift}$, but is a reasonable approximation for the scattering processes involving the momentum transfer [24]. We apply the Markov approximation to the scattering part of Eq. (7) which therefore rewrites:

$$\frac{d\rho}{dt} = \frac{1}{i\hbar}\left[H_{shift}(t),\rho\right] - \frac{1}{\hbar^2}\int_{-\infty}^{t}\left[H_{scatt}(t),\left[H_{scatt}(\tau),\rho(\tau)\right]\right]d\tau . \tag{8}$$

The next step is to factorize the density matrix of the system as a product of the phonon density matrix and polariton density matrix corresponding to the different states in the reciprocal space

$$\rho = \rho_{ph}\otimes\prod_{\mathbf{k}}\rho_{\mathbf{k}} \tag{9}$$

(Born approximation). The phonons are then "traced out", i.e. in the calculation their occupation numbers are taken as fixed parameters determined by the temperature.

The equations (8) and (9) can be rewritten in terms of physically observable quantities, such as occupation numbers of the spin-up and spin-down polaritons and the in-plane components of the pseudospins [25]

$$N_{\mathbf{k}\uparrow} = Tr\left(p_{\mathbf{k}\uparrow}^{+} p_{\mathbf{k}\uparrow} \rho\right),$$
$$N_{\mathbf{k}\downarrow} = Tr\left(p_{\mathbf{k}\downarrow}^{+} p_{\mathbf{k}\downarrow} \rho\right),$$
$$S_{\mathbf{k},x} + iS_{\mathbf{k},y} = Tr\left(p_{\mathbf{k}\downarrow}^{+} p_{\mathbf{k}\uparrow} \rho\right)$$

(10)

Z-component of the pseudospin is given by $S_{\mathbf{k},z} = \left(N_{\mathbf{k}\uparrow} - N_{\mathbf{k}\downarrow}\right)/2$. We remind, that the pseudospin describes both the exciton spin state and its dipole moment orientation. The polaritons with a pseudospin oriented along Z-axis emit circularly polarized light. Their linear combinations correspond to eigenstates of $S_x$ and $S_y$ yielding linearly polarized emission. The pseudospin parallel to X-axis corresponds to X-polarized light (exciton), the pseudospin antiparallel to X-axis corresponds to Y-polarized light, the pseudospin oriented along Y-axis describes diagonal linear polarizations.

Once polariton populations and pseudospins are known, the intensities of the circular and linear components of photoemission can be found from

$$I_{\mathbf{k}}^{+} = N_{\uparrow,\mathbf{k}}$$
$$I_{\mathbf{k}}^{-} = N_{\downarrow,\mathbf{k}}$$
$$I_{\mathbf{k}}^{x} = \frac{N_{\uparrow,\mathbf{k}} + N_{\downarrow,\mathbf{k}}}{2} + S_{x,\mathbf{k}}$$
$$I_{\mathbf{k}}^{y} = \frac{N_{\uparrow,\mathbf{k}} + N_{\downarrow,\mathbf{k}}}{2} - S_{x,\mathbf{k}}$$

(11)

Note that polarizations of light parallel to X and Y axes correspond to the pseudospin parallel and antiparallel to X axis, respectively.

The general form of the kinetic equations written in terms of the occupation numbers and pseudospins reads:

$$\frac{dN_{\mathbf{k}\uparrow,\downarrow}}{dt} = -\frac{N_{\mathbf{k}\uparrow,\downarrow}}{\tau_k} + \left(\frac{dN_{\mathbf{k}\uparrow,\downarrow}}{dt}\right)\bigg|_{rot} + \left(\frac{dN_{\mathbf{k}\uparrow,\downarrow}}{dt}\right)\bigg|_{phon} + \left(\frac{dN_{\mathbf{k}\uparrow,\downarrow}}{dt}\right)\bigg|_{pol-pol}$$

(12)

$$\frac{d\mathbf{S}_{\perp,\mathbf{k}}}{dt} = -\frac{\mathbf{S}_{\perp,\mathbf{k}}}{\tau_k} + \left(\frac{d\mathbf{S}_{\perp,\mathbf{k}}}{dt}\right)\bigg|_{rot} + \left(\frac{d\mathbf{S}_{\perp,\mathbf{k}}}{dt}\right)\bigg|_{phon} + \left(\frac{d\mathbf{S}_{\perp,\mathbf{k}}}{dt}\right)\bigg|_{pol-pol}$$

(13)

where the first term describes the radiative decay of exciton polaritons, the indices *rot*, *phon*, *pol-pol* correspond to the pseudospin rotation in the effective magnetic field, the scattering with acoustic phonons and polariton- polariton scattering respectively.

The rotation terms read:

$$\left(\frac{dN_{\mathbf{k}\uparrow}}{dt}\right)\bigg|_{rot} = -\left(\frac{dN_{\mathbf{k}\downarrow}}{dt}\right)\bigg|_{rot} = \mathbf{e}_z \cdot \left[\mathbf{S}_{\perp,\mathbf{k}} \times \mathbf{\Omega}_{LT,\mathbf{k}}\right] \qquad , \tag{14}$$

$$\left(\frac{d\mathbf{S}_{\perp,\mathbf{k}}}{dt}\right)\bigg|_{rot} = \left[\mathbf{S}_{\perp,\mathbf{k}} \times \mathbf{\Omega}_{int,\mathbf{k}}\right] + \frac{\left(N_{\mathbf{k}\uparrow} - N_{\mathbf{k}\downarrow}\right)}{2}\bar{\mathbf{\Omega}}_{LT,\mathbf{k}} \qquad , \tag{15}$$

where $\mathbf{e}_z$ is a unitary vector in the direction of the structure growth axis, $\mathbf{\Omega}_{LT,\mathbf{k}}$ is an effective in- plane magnetic field produced by TE- TM splitting, $\bar{\mathbf{\Omega}}_{LT,\mathbf{k}}$ is obtained from $\mathbf{\Omega}_{LT,\mathbf{k}}$ by the rotation by 90° about the structure growth axis, the effective magnetic field $\mathbf{\Omega}_{int,\mathbf{k}}$ produced by the imbalance of the $\sigma^+$ and $\sigma^-$ polaritons is given by the following expression deduced from $H_{shift}$:

$$\hbar\mathbf{\Omega}_{int,\mathbf{k}} = 2\mathbf{e}_z \sum_{\mathbf{k'}}\left(V^{(1)}_{\mathbf{k},\mathbf{k'},0} - V^{(2)}_{\mathbf{k},\mathbf{k'},0}\right)\left(N_{\mathbf{k'}\uparrow} - N_{\mathbf{k'}\downarrow}\right) \quad . \tag{16}$$

The dynamics with acoustic phonons is given by the following expression :

$$\left(\frac{dN_{\mathbf{k}\uparrow,\downarrow}}{dt}\right)\bigg|_{phon} = \sum_{\mathbf{k'}}\left[\left(W_{\mathbf{k'k}} - W_{\mathbf{kk'}}\right)\left(N_{\mathbf{k}\uparrow,\downarrow}N_{\mathbf{k'}\uparrow,\downarrow} + (\mathbf{S}_{\mathbf{k}} \cdot \mathbf{S}_{\mathbf{k'}})\right) + \left(W_{\mathbf{k'k}}N_{\mathbf{k'}\uparrow,\downarrow} - W_{\mathbf{kk'}}N_{\mathbf{k}\uparrow,\downarrow}\right)\right] \quad , \tag{17}$$

$$\left(\frac{d\mathbf{S}_{\perp,\mathbf{k}}}{dt}\right)\bigg|_{phon} = \sum_{\mathbf{k'}}\left[\frac{1}{2}\left(W_{\mathbf{k'k}} - W_{\mathbf{kk'}}\right)\left(\left(N_{\mathbf{k}\uparrow} + N_{\mathbf{k}\downarrow}\right)\mathbf{S}_{\perp,\mathbf{k'}} + \left(N_{\mathbf{k'}\uparrow} + N_{\mathbf{k'}\downarrow}\right)\mathbf{S}_{\perp,\mathbf{k}}\right) + \left(W_{\mathbf{k'k}}\mathbf{S}_{\perp,\mathbf{k'}} - W_{\mathbf{kk'}}\mathbf{S}_{\perp,\mathbf{k}}\right)\right] \tag{18}$$

where the scattering rates are

$$W_{\mathbf{kk'}} = \begin{cases} \dfrac{2\pi}{\hbar}\left|U_{\mathbf{k},\mathbf{k'-k}}\right|^2 n_{ph,\mathbf{k'-k}}\delta\left(E(\mathbf{k'}) - E(\mathbf{k}) - \varepsilon_{phon,\mathbf{k'-k}}\right), |\mathbf{k'}| > |\mathbf{k}| \\[2mm] \dfrac{2\pi}{\hbar}\left|U_{\mathbf{k},\mathbf{k'-k}}\right|^2\left(n_{ph,\mathbf{k'-k}} + 1\right)\delta\left(E(\mathbf{k'}) - E(\mathbf{k}) + \varepsilon_{phon,\mathbf{k'-k}}\right), |\mathbf{k'}| < |\mathbf{k}| \end{cases} \tag{19}$$

The Dirac delta functions in (19) account for the energy conservation during any scattering act. Mathematically they appear from integration of the time-dependent exponents in the second term of Eq. (8). Writing energy delta functions in (19) we suppose that the polariton TE-TM splitting is not strong enough with respect to the broadening of the states due to their finite radiative life-time. Thus the TE-TM splitting does not modify sufficiently the polariton dispersion curve, and it can be neglected while calculating $E(\mathbf{k'}), E(\mathbf{k})$. In numerical calculations, the delta functions should be replaced by resonant functions having finite amplitudes proportional to the inverse energy broadenings of corresponding polariton states.

The part describing the polariton-polariton interactions reads

$$\left(\frac{dN_{\mathbf{k}\uparrow}}{dt}\right)\bigg|_{pol-pol} = \sum_{\mathbf{k',q}} \Big\{ W^{(1)}_{\mathbf{k,k',q}} \Big[ (N_{\mathbf{k}\uparrow} + N_{\mathbf{k'}\uparrow} + 1) N_{\mathbf{k+q}\uparrow} N_{\mathbf{k'-q}\uparrow} - (N_{\mathbf{k+q}\uparrow} + N_{\mathbf{k'-q}\uparrow} + 1) N_{\mathbf{k}\uparrow} N_{\mathbf{k'}\uparrow} \Big] +$$

$$+ W^{(2)}_{\mathbf{k,k',q}} \Big[ (N_{\mathbf{k}\uparrow} + N_{\mathbf{k'}\downarrow} + 1)(N_{\mathbf{k+q}\uparrow} N_{\mathbf{k'-q}\downarrow} + N_{\mathbf{k+q}\downarrow} N_{\mathbf{k'-q}\uparrow} + 2(\mathbf{S_{\perp k+q} \cdot S_{\perp k'-q}})) -$$

$$(N_{\mathbf{k}\uparrow} N_{\mathbf{k'}\downarrow} + (\mathbf{S_{\perp k} \cdot S_{\perp k'}}))(N_{\mathbf{k+q}\uparrow} + N_{\mathbf{k'-q}\downarrow} + N_{\mathbf{k+q}\downarrow} + N_{\mathbf{k'-q}\uparrow} + 2) \Big] +$$

$$+ 2W^{(2)}_{\mathbf{k,k'}\uparrow} \Big[ N_{\mathbf{k'}\uparrow} (\mathbf{S_{\perp k'} \cdot S_{\perp k'-q}}) + N_{\mathbf{k'-q}\uparrow} (\mathbf{S_{\perp k'} \cdot S_{\perp k+q}}) - N_{\mathbf{k}\uparrow} \mathbf{S_{\perp k'}} \cdot (\mathbf{S_{\perp k'-q}} + \mathbf{S_{\perp k+q}}) \Big] +$$

$$+ W^{(12)}_{\mathbf{k,k',q}} [(\mathbf{S_{\perp k} \cdot S_{\perp k+q}})(N_{\mathbf{k'-q}\uparrow} + N_{\mathbf{k'-q}\downarrow} - N_{\mathbf{k}\uparrow} - N_{\mathbf{k'}\downarrow}) + (\mathbf{S_{\perp k} \cdot S_{\perp k'-q}})(N_{\mathbf{k+q}\uparrow} + N_{\mathbf{k+q}\downarrow} - N_{\mathbf{k}\uparrow} - N_{\mathbf{k'}\downarrow})] \Big\}$$

$$(20)$$

$$\left(\frac{d\mathbf{S_{\perp k}}}{dt}\right)\bigg|_{pol-pol} = \sum_{\mathbf{k',q}} \Bigg\{ \frac{W^{(1)}_{\mathbf{k,k',q}}}{2} \mathbf{S_{\perp k}} \Big[ N_{\mathbf{k+q}\uparrow} N_{\mathbf{k'-q}\uparrow} + N_{\mathbf{k+q}\downarrow} N_{\mathbf{k'-q}\downarrow} - N_{\mathbf{k'}\uparrow}(N_{\mathbf{k+q}\uparrow} + N_{\mathbf{k'-q}\uparrow} + 1) - N_{\mathbf{k'}\downarrow}(N_{\mathbf{k+q}\downarrow} + N_{\mathbf{k'-q}\downarrow} + 1) \Big] +$$

$$+ W^{(1)}_{\mathbf{k,k',q}} \Big( \mathbf{S_{\perp k+q}}(\mathbf{S_{\perp k'} \cdot S_{\perp k'-q}}) + \mathbf{S_{\perp k'-q}}(\mathbf{S_{\perp k'} \cdot S_{\perp k+q}}) - \mathbf{S_{\perp k'}}(\mathbf{S_{\perp k+q} \cdot S_{\perp k'-q}}) \Big) +$$

$$+ \frac{W^{(2)}_{\mathbf{k,k',q}}}{2} \Big[ 2(\mathbf{S_{\perp k}} + \mathbf{S_{\perp k'}}) \Big( N_{\mathbf{k+q}\uparrow} N_{\mathbf{k'-q}\downarrow} + N_{\mathbf{k+q}\downarrow} N_{\mathbf{k'-q}\uparrow} + 2(\mathbf{S_{\perp k+q} \cdot S_{\perp k'-q}}) \Big) -$$

$$(\mathbf{S_{\perp k}}(N_{\mathbf{k'}\uparrow} + N_{\mathbf{k'}\downarrow}) + \mathbf{S_{\perp l}}(N_{\mathbf{k}\uparrow} + N_{\mathbf{k}\downarrow}))(N_{\mathbf{k+q}\uparrow} + N_{\mathbf{k'-q}\uparrow} + N_{\mathbf{k+q}\downarrow} + N_{\mathbf{k'-q}\downarrow} + 2) \Big] -$$

$$- 2W^{(12)}_{\mathbf{k,k',q}} \vec{S}_{\perp \mathbf{k}} \Big( (\mathbf{S_{\perp k'} \cdot S_{\perp k+q}}) + (\mathbf{S_{\perp k'} \cdot S_{\perp k'-q}}) \Big) +$$

$$+ \frac{W^{(12)}_{\mathbf{k,k',q}}}{2} \mathbf{S_{\perp k'-q}} \Big[ 2((N_{\mathbf{k'}\uparrow} + 1)N_{\mathbf{k+q}\uparrow} + (N_{\mathbf{k'}\downarrow} + 1)N_{\mathbf{k+q}\downarrow}) + (N_{\mathbf{k+q}\uparrow} + N_{\mathbf{k+q}\downarrow} - N_{\mathbf{k'}\uparrow} - N_{\mathbf{k'}\downarrow})(N_{\mathbf{k}\uparrow} + N_{\mathbf{k}\downarrow}) \Big] +$$

$$+ \frac{W^{(12)}_{\mathbf{k,k',q}}}{2} \mathbf{S_{\perp k+q}} \Big[ 2((N_{\mathbf{k'}\uparrow} + 1)N_{\mathbf{k'-q}\uparrow} + (N_{\mathbf{k'}\downarrow} + 1)N_{\mathbf{k'-q}\downarrow}) + (N_{\mathbf{k'-q}\uparrow} + N_{\mathbf{k'-q}\downarrow} - N_{\mathbf{k'}\uparrow} - N_{\mathbf{k'}\downarrow})(N_{\mathbf{k}\uparrow} + N_{\mathbf{k}\downarrow}) \Big] \Bigg\}$$

$$(21)$$

(dans cette equation $-2W^{(12)}_{\mathbf{k,k',q}}\mathbf{S_{\perp k}}$)

The equation for $N_{\mathbf{k}\downarrow}$ can be obtained from (20) by changing the spin index. The scattering rates are defined as follows

$$W^{(1)}_{\mathbf{k,k',q}} = \frac{2\pi}{\hbar} \left| V^{(1)}_{\mathbf{k,k',q}} \right|^2 \delta\big(E(\mathbf{k}) + E(\mathbf{k'}) - E(\mathbf{k+q}) - E(\mathbf{k'-q})\big)$$

$$W^{(2)}_{\mathbf{k,k',q}} = \frac{2\pi}{\hbar} \left| V^{(2)}_{\mathbf{k,k',q}} \right|^2 \delta\big(E(\mathbf{k}) + E(\mathbf{k'}) - E(\mathbf{k+q}) - E(\mathbf{k'-q})\big)$$

$$(22)$$

$$W^{(12)}_{\mathbf{k,k',q}} = \frac{2\pi}{\hbar} \mathrm{Re}\big(V^{(1)}_{\mathbf{k,k',q}} V^{*(2)}_{\mathbf{k,k',q}}\big) \delta\big(E(\mathbf{k}) + E(\mathbf{k'}) - E(\mathbf{k+q}) - E(\mathbf{k'-q})\big)$$

Here as in Eq. (19) the delta functions ensure the energy conservation. The signs of the terms $W^{(1)}_{\mathbf{k,k',q}}, W^{(2)}_{\mathbf{k,k',q}}$ and $W^{(12)}_{\mathbf{k,k',q}}$ can differ in different systems. Although $W^{(1)}_{\mathbf{k,k',q}}, W^{(2)}_{\mathbf{k,k',q}}$ are always positive, $W^{(12)}_{\mathbf{k,k',q}}$ can be positive or negative depending on the phase shift between the matrix elements of the singlet and triplet scattering. In particular it is negative if these matrix elements are real and have opposite signs. As one can see, the system of equations (20,21) is rather heavy for analysis. The corresponding numerical simulations also can hardly be carried out without sufficient simplifications. To make the simulations feasible one can either neglect polariton-polariton collisions and consider the scattering with acoustic phonons and free electrons as the only mechanism of the energy relaxation in the system [13], or retain all the scattering mechanisms but restrict the number of states in the reciprocal space to some reasonable minimum. The latter approach is standard in description of the polariton parametric amplifier. Usually only three states are taken into account: the signal, the pump and the idler [7,8]. This is a reasonable approximation in the cavities with strong negative detuning where these three states have extremely small wave numbers and thus lie at the photonic part of the LPB, so that their scattering with acoustic phonons is prohibited by energy and momentum conservation. On the other hand, for zero and positive detuning the pump and idler states are efficiently coupled with a thermal reservoir of polaritons in the excitonic part of the LPB. As we show in this paper, this coupling should be taken into account in order to achieve a satisfactory description of the influence of the detuning on the spin dynamics of parametric amplifiers.

Figure 1 shows a schematic diagram of the polariton parametric amplifier. The LPB dispersion curve has been calculated based on the parameters of the samples used in experiments. Our approach is to treat the polariton parametric amplifier as a spinor three-level system (pump, signal and idler) connected with a dissipative reservoir which consists mainly of bare exciton states, very close in energy to the pump and idler states in the zero detuning case. This coupling provides an additional mechanism of decay for the polariton population and pseudo-spin. Pump, signal and idler are coupled by the polariton- polariton parametric scattering, while the reservoir is taken into account by introduction of an effective spin decay time of the three main states. Polariton-polariton scattering is treated within the framework of the Born-Markov approximation (for the description of the spinor parametric amplifier beyond markovian approximation see Ref. [26]).

We solve numerically the following system of kinetic equations:

$$\frac{dN_{s\uparrow}}{dt} = -\frac{N_{s\uparrow}}{\tau_s} + \left(\frac{dN_{s\uparrow}}{dt}\right)\bigg|_{po} - N_{s\uparrow}\left(\frac{1}{\tau_s^r} + \frac{1}{2\tau_{s,z}^r}\right) + \frac{N_{s\downarrow}}{2\tau_{s,z}^r}, \qquad (23)$$

$$\frac{dN_{s\downarrow}}{dt} = -\frac{N_{s\downarrow}}{\tau_s} + \left(\frac{dN_{s\downarrow}}{dt}\right)\bigg|_{po} - N_{s\downarrow}\left(\frac{1}{\tau_s^r} + \frac{1}{2\tau_{s,z}^r}\right) + \frac{N_{s\uparrow}}{2\tau_{s,z}^r}, \qquad (23a)$$

$$\frac{d\mathbf{S}_{s\perp}}{dt} = -\mathbf{S}_{s\perp}\left(\frac{1}{\tau_s} + \frac{1}{\tau_s^r}\right) + \left(\frac{d\mathbf{S}_{s\perp}}{dt}\right)\bigg|_{po} + [\mathbf{S}_{s\perp}\times\mathbf{\Omega}_{\text{int}}] - \frac{S_s^x}{\tau_{s,x}^r} - \frac{S_s^y}{\tau_{s,y}^r}, \qquad (23b)$$

$$\frac{dN_{p\uparrow}}{dt} = -\frac{N_{p\uparrow}}{\tau_p} + \left(\frac{dN_{p\uparrow}}{dt}\right)\bigg|_{po} - N_{p\uparrow}\left(\frac{1}{\tau_p^r} + \frac{1}{2\tau_{p,z}^r}\right) + \frac{N_{p\downarrow}}{2\tau_{p,z}^r}, \qquad (23c)$$

$$\frac{dN_{p\downarrow}}{dt} = -\frac{N_{p\downarrow}}{\tau_p} + \left(\frac{dN_{p\downarrow}}{dt}\right)\bigg|_{po} - N_{p\downarrow}\left(\frac{1}{\tau_p^r} + \frac{1}{2\tau_{p,z}^r}\right) + \frac{N_{p\uparrow}}{2\tau_{p,z}^r}, \qquad (23d)$$

$$\frac{d\mathbf{S}_{p\perp}}{dt} = -\mathbf{S}_{p\perp}\left(\frac{1}{\tau_p} + \frac{1}{\tau_p^r}\right) + \left(\frac{d\mathbf{S}_{p\perp}}{dt}\right)\bigg|_{po} + [\mathbf{S}_{p\perp}\times\mathbf{\Omega}_{\text{int}}] - \frac{S_p^x}{\tau_{p,x}^r} - \frac{S_p^y}{\tau_{p,y}^r}, \qquad (23e)$$

$$\frac{dN_{i\uparrow}}{dt} = -\frac{N_{i\uparrow}}{\tau_i} + \left(\frac{dN_{i\uparrow}}{dt}\right)\bigg|_{po} - N_{i\uparrow}\left(\frac{1}{\tau_i^r} + \frac{1}{2\tau_{i,z}^r}\right) + \frac{N_{i\uparrow}}{2\tau_{i,z}^r} \ , \tag{23f}$$

$$\frac{dN_{i\downarrow}}{dt} = -\frac{N_{i\downarrow}}{\tau_i} + \left(\frac{dN_{i\downarrow}}{dt}\right)\bigg|_{po} - N_{i\downarrow}\left(\frac{1}{\tau_i^r} + \frac{1}{2\tau_{i,z}^r}\right) + \frac{N_{i\downarrow}}{2\tau_{i,z}^r} \ , \tag{23g}$$

$$\frac{d\mathbf{S}_{i\perp}}{dt} = -\mathbf{S}_{i\perp}\left(\frac{1}{\tau_i} + \frac{1}{\tau_i^r}\right) + \left(\frac{d\mathbf{S}_{i\perp}}{dt}\right)\bigg|_{po} + [\mathbf{S}_{i\perp} \times \mathbf{\Omega}_{\mathrm{int}}] - \frac{\mathbf{S}_i^x}{\tau_{i,x}^r} - \frac{\mathbf{S}_i^y}{\tau_{i,y}^r} \ , \tag{23h}$$

where the indices *s,p,* and *i* correspond to the signal, pump, and idler respectively. The index *po* corresponds to the parametric process described by Eqs. (20) and (21) (where the summation over **q** and **k'** should be omitted in this case). $\tau_{s,p,i}$ are the radiative lifetimes of signal, pump and idler given by $\tau_{s,p,i} = \dfrac{\tau_0}{\left|C_{L;k_s,k_p,k_i}\right|^2}$

where $\tau_0$ is the photon life time in the cavity and $\left|C_{L;k_s,k_p,k_i}\right|^2$ are the photon fractions of signal, pump and idler respectively. The $\tau_{s,p,i}^r$ are the decay times associated with irreversible escape of particles to the reservoir (these are much longer than the radiative lifetimes), and $\tau_{s,p,i;x,y,z}^r$ are the spin relaxation times induced by the coupling to the reservoir of the x,y,z components of the pseudospin of signal, pump and idler respectively. The coupling of a state with the reservoir depends on its exciton fraction, on temperature (via the occupation of acoustic phonons modes) and on the energy difference between the state and the reservoir. These decay times are therefore given by

$$\tau_{s,p,i,x,y,z}^r = \tau_{x,y,z}^r \left|X_{L;k_s,k_p,k_i}\right|^2 e^{-\left(\frac{E_r - E_{s,p,i}}{k_B T}\right)} \tag{24}$$

where $\left|X_{L;k_s,k_p,k_i}\right|^2$ are the exciton fractions in signal pump and idler, $E_{r,s,p,i}$ are the energies of the reservoir, signal, pump, and idler, $\tau_{x,y,z}^r$ are the fitting parameters related to the exciton pseudospin decay time in the reservoir. Note that the signal is always effectively separated from the reservoir because of the large energy difference.

### 3. Results and discussion

#### 3-a) Samples description and experimental setup

The sample 1 consists of two Bragg mirrors made of 22(26) *AlAs/Al$_{0.1}$Ga$_{0.9}$ As* layers with a single *8 nm* wide *Ga$_{0.95}$In$_{0.05}$ As* QW embedded in the middle. The vacuum Rabi splitting is 3.5 *meV* and the photon lifetime in the cavity is $\tau_c \sim 8~ps$. The sample 2 is similar to the previous one. It consists of two Bragg mirrors made of 17(27) *AlAs/Al$_{0.1}$Ga$_{0.9}$ As* layers; a single *8 nm* wide *Ga$_{0.95}$In$_{0.05}$ As* quantum well (QW) is embedded in the middle. The vacuum Rabi splitting is 3.7 *meV*, and the photon lifetime in the cavity is $\tau_c \sim 3~ps$. In both samples, the cavity is wedged, so that the detuning $\bar{\delta}$ between the cavity and exciton modes could be varied by moving the excitation spot on the sample surface.

We apply the theoretical formalism developed in the previous Section for description of the experimental results on the polarization dynamics of photoluminescence of two λ-microcavities. Both samples display the same qualitative behaviour. We show the results only for sample 1 in the following.

The excitation beam, resonant with the LPB, is incident at an angle of 8° (±1°), so that it generates LPB polaritons in a state with initial in-plane wave vector $kp \approx 1 \times 10^4$ cm$^{-1}$ . The excitation spot diameter is about 100 μm. Its spectral width is about 1.5 meV. The ellipticity of the incident light is tuned using a Soleil-Babinet compensator. The polariton emission is recorded at 10K using a two-colour up-conversion spectroscopy set-up, which provides time and spectral resolutions of about 2 ps, and 1.5 meV respectively [27]. In particular, the cavity photon lifetime is deduced from time resolved emission measurements performed at negative cavity detuning ($\bar{\delta}$ = - 9 meV for sample 1 and $\bar{\delta}$ = - 3 meV for sample 2). We have characterised experimentally the detection solid angle determined by the frequency-mixing process: it consists in an elliptical cone, normal to the micro-cavity surface, with aperture half angles $\theta$x ≈ 3.8° and $\theta$y ≈ 0.95° in the horizontal plane (containing $\bm{kp}$) and vertical plane (orthogonal to $\bm{kp}$) respectively. The resulting small acceptance solid angle $\delta\Omega$ ($\delta\Omega \approx 3 \times 10^{-3}$ steradian) allows us to detect the polariton modes in the vicinity of k ≈ 0. In the linear excitation case the polarization of the incident light was parallel to the crystallographic axis [110].

#### 3-b) Parameters of calculation

As explained in the Section 2 and illustrated by figure 1, we have chosen a model accounting for three discrete polariton states, namely the signal state, the pump state, and the idler state. The excitonic reservoir is located in the vicinity of the bare exciton energy and is characterized by spin relaxation times $\tau_{x,y,z}^r$. As the wave vectors corresponding to signal, pump and idler states are relatively small, the TE-TM-splitting of polariton modes is also small there, ranging from 0 at signal to a few $\mu$ eV at the idler state, which yields the pseudospin rotation period of about 300 ps in the sample we consider. The field $\mathbf{\Omega}_{\mathrm{int}}$ describes the effects of the self-induced Larmor precession and depends on the polariton concentration and on the circular polarization degree of pump. In the reservoir, the TE-TM splitting can be accounted for by introduction of an effective magnetic field randomly oriented in the plane of the structure. In this case, the in-plane linear polarization should decay twice slower than circular one in the reservoir. In the motional narrowing regime we arrive to

$$\tau_{rs}^{circ} = \frac{1}{\left\langle \Omega_{LT,r}^2 \right\rangle \tau_{scatt}}, \quad \tau_{rs}^{lin} = \frac{2}{\left\langle \Omega_{LT,r}^2 \right\rangle \tau_{scatt}} \quad , \tag{25}$$

where $\hbar \left\langle \Omega_{LT,r}^2 \right\rangle^{1/2}$ is an average value of the LT-splitting in the reservoir, $\tau_{scatt}$ is a characteristic scattering time within the reservoir, $\tau_{rs}^{circ}$ is the decay time of the circular polarization and $\tau_{rs}^{lin}$ is that the decay time of linear polarization. As we shall see in the next section, this picture does not fit well with our experimental findings which show a relatively fast relaxation of the circular polarization in the reservoir, and almost no relaxation of the linear polarization. This suggests that in fact the relaxation times in two polarizations differ much stronger than what is predicted by Eq. (25). As clearly follows from the experimental data, the polarization parallel to the crystallographic axes [110] and [1-10] is quite well conserved in the reservoir. The most likely explanation of it would be that the reservoir is mainly composed of localized exciton states which are known to be strongly polarized along the main crystallographic axes [28]. We shall use this hypothesis in the further analysis. Choosing the X-axis in pseudo-spin space parallel to the [110] and [1-10], we shall assume: $\tau_y^r = \tau_z^r = 11\,ps << \tau_x^r = 200\,ps$. These two times are the only adjustable parameters of the model.

To calculate the matrix element of exciton-exciton interaction, we have used the formula $V_{k,k',q}^{ex(1)} = 6E_b \dfrac{a_b^2}{S}$ where $E_b$ is the exciton binding energy, $a_b$ the 2 dimensional Bohr radius and $S$ the laser spot size [17].

### 3-c) Detuning dependence of the spin relaxation rates

Figure 2 shows the experimental data and theoretical simulation for sample 1 at different detunings in the spontaneous scattering regime. At negative detuning the system exhibits a polarization lifetime of about 300 ps as well as fast decay of the emission (30 ps). This fast decay shows that polaritons are not scattered out of the light cone (polariton trap zone) before to escape from the cavity by tunneling across the mirrors. Three states composing the parametric amplifier have energies deeply inside the polariton trap and are therefore efficiently protected from the influence of the reservoir. The polarisation decay rate for these states approaches the intrinsic one, given by the value of the TE-TM splitting in the pump state.

At zero and positive detuning the situation changes drastically. In this cavity and at these detunings, the energy spacing between the pump and the reservoir becomes smaller than 1.5 meV which is small enough to allow for their efficient coupling by phonon or free carrier scattering. As a result, a much longer decay time of the emission is observed. After a short initial polarization plateau (t<10 ps) which we attribute to surface Rayleigh scattering, the circular polarization drops with a decay time of about 100 ps. At short time delay, the ground state emission is governed by parametric luminescence mechanism. Later (t>30-50 ps), the particle remaining in the system lose their polarization due to interaction with the reservoir. Simultaneously, the polarisation shows a small plateau of about 10 ps before to drop with a decay time of about 100 ps. This shows that in the first 10 ps, the ground state emission is governed by the parametric luminescence mechanism. After that, most of the particles remaining in the system have lost their polarization by interacting with the reservoir.

### 3-d) Inversion of the linear polarisation degree.

Figure 3 shows the experimental data and corresponding theoretical simulation for the sample 1 under linear pumping, below and above the stimulation threshold ($\delta=0$). Below threshold, the polarisation of emission is almost constant (-8%) and is opposite to the pump polarisation. Above threshold, the polarisation degree achieves a large negative value of −65 % and then, in the long time limit, decreases to the spontaneous scattering value of −8 %. These data are very well reproduced theoretically (see figure 3(b)). This peculiar non-monotonic behaviour of the linear polarization degree of emission is due to the anisotropy of the polariton-polariton interaction, as we are going to show now. In the spontaneous regime, Eq. (23) which describes the motion of the signal pseudo-spin can be simplified and reduced to:

$$\frac{d\mathbf{S}_{s,\perp}}{dt} = W^{(12)}_{0,k_p,k_p}\left(N_{\uparrow_p} + N_{p\downarrow}\right)\mathbf{S}_{p,\perp} - \frac{\mathbf{S}_{s,\perp}}{\tau_s} \quad, \tag{26}.$$

where $k_p$ is the in-plane wave vector of the pump. An analytical formula can be written for the linear polarisation degree of the signal in this case:

$$\rho_l = \frac{4W^{(12)}_{0,k_p,k_p}}{W^{(1)}_{0,k_p,k_p} + 4W^{(2)}_{0,k_p,k_p}} \quad. \tag{27}$$

Note that the similar formula has been first proposed in Ref. [18] from qualitative arguments. In the isotropic case ($2V^{(2)}_{0,k_p,k_p} = V^{(1)}_{0,k_p,k_p}$) the linear polarisation degree of the signal should be +100 %. If $V^{(2)}$ is neglected, the memory of the initial linear polarisation is lost and $\rho_l$ is equal to 0 as one can also see from Ref. [29]. For our microcavity sample, the relative values of scattering constants have been estimated as [18]:

$$2V_{0,k_p,k_p}^{(2)} = -0.08V_{0,k_p,k_p}^{(1)}, \qquad (28)$$

$V_{0,k_p,k_p}^{(1)}$ and $V_{0,k_p,k_p}^{(2)}$ have opposite signs which reflects the different mechanisms of interaction between polaritons having the parallel spins (triplet configuration) and opposite spins (singlet configuration). In triplet configuration the interaction is repulsive because of the Pauli principle. On the other hand, the singlet configuration can lead to formation of an excitonic molecule (bi-exciton or bi-polariton), thus interaction between polaritons having opposite spins is likely to be attractive. Our measurements and analysis show that polariton-polariton scattering is a very efficient channel of relaxation of the linear polarisation in microcavities. Actually, in the spontaneous regime 92% of initial linear polarisation may be lost in a single polariton-polariton scattering event. On the other hand, in the stimulated regime, 8% of remaining linear polarization constitute a seed which allows to build up a huge population of polarized polaritons and achieve the negative polarisation degree of about 65 % .

*3-e) Direct observation of the self-induced Larmor precession in the time domain.*

The self-induced Larmor precession [16] is one of the effects caused by the anisotropy of the polariton-polariton interaction (i.e. dependence of the scattering amplitude on mutual orientation of the pseudospins of two interacting polaritons). Because of this anisotropy, an imbalance between $\sigma^+$ and $\sigma^-$ populations is responsible for the appearance of an effective magnetic field in the Z-direction (see Eq. (16)) which is able to rotate the linear polarisation plane. Figure 4 shows time dependences of circular and linear polarization degrees for elliptic pumping at different pumping intensities (below and above threshold) measured on the sample 1. Below the threshold, the circular polarisation degree decays similarly to the circular excitation case. On the other hand, the linear polarization shows a single oscillation which starts from negative values of the polarization degree. We interpret this oscillation as a manifestation of the self-induced Larmor precession effect. Above the stimulation threshold, the circular polarisation degree of the emission is again very similar to the circular excitation case showing a long plateau followed by a fast decay. The oscillation of the linear polarisation degree is still visible, but remarkably it has a longer period than in the spontaneous case both in experiment and theory. This is easy to understand, as the effective magnetic field acting on the polariton pseudospin is density dependent and its intensity decreases as time goes. Thus the oscillation period should rapidly increase versus time in a pulsed experiment. In the stimulated regime, the circular polarisation degree is kept for a long time, but the intensity of emission decays much faster. Thus the overall effective field which rotates the in-plane polarisation is decaying faster in the stimulated regime than in the spontaneous regime. That is why the period self-induced Larmor precession is much longer if the stimulated scattering dominates.

**4. Conclusions**

In conclusion, we have derived a set of kinetic equations describing energy and spin relaxation of exciton polaritons in the framework of the Born-Markov approximation. All important mechanisms of the polariton redistribution in the reciprocal space including polariton- polariton scattering have been taken into account. The spin-dependence of polariton- polariton interactions results in a number of peculiar phenomena, including self-induced Larmor precession and linear polarisation inversion.

This model has been applied to describe the polarization dynamics of photoluminescence from resonantly pumped GaAs microcavities having a single InGaAs/GaAs quantum well embedded. For the circular pump, we demonstrate that interaction with reservoir states has a crucial influence on the decay time of the circular polarization of polariton emission. The coupling to the reservoir depends on the detuning parameter and temperature, which also makes the spin relaxation time of the parametric luminescence detuning dependent. For linear pump, the observed inversion of the degree of linear polarisation of signal emission has been interpreted as a result of strongly different amplitudes of polariton-polariton scattering in the singlet and triplet configurations. The increase of the absolute value of the linear polarization degree of the signal with pump intensity has been explained in terms of the spin-dependent final state bosonic stimulation of polariton-polariton scattering. Long decay time of linear polarization has been attributed to the anisotropy of the excitonic states in the reservoir, polarized preferentially along [110] and [1-10] crystallographic axes. The oscillations of the linear polarization in the case of elliptically polarized pump have been interpreted as a result of the self-induced Larmor precession.

We thank Dr. K. Kavokin for useful discussion we had on the topic. This work has been supported by the Marie-Curie RTN "Clermont2" (contract MRTN-CT-2003-503677) and by the "ACI Polariton" project. AK and GM acknowledge support from the ARL- ERO project number N62558-06-P0095. MMG is grateful to RFBR and "Dynasty" foundation - ICFPM for financial support.

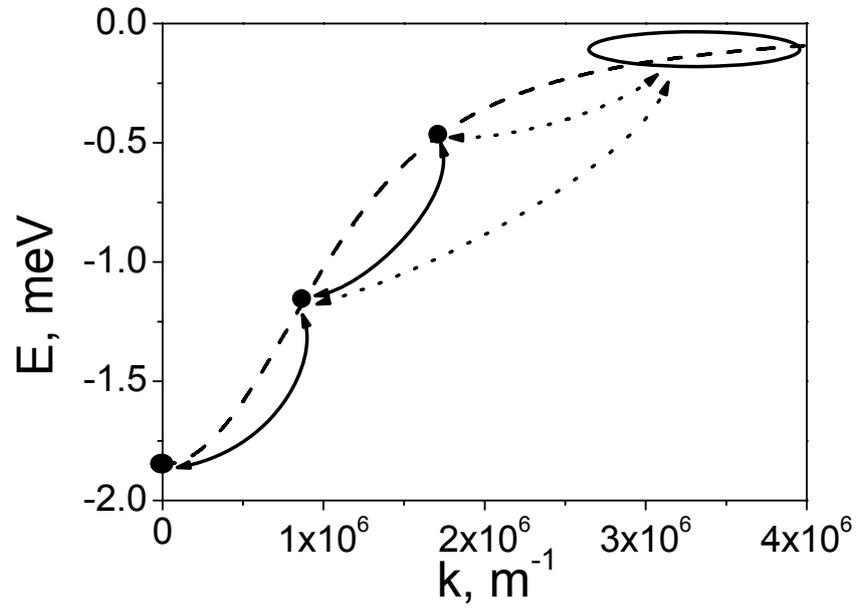

Figure 1. Schematic diagram of parametric amplifier with reservoir. The dashed line is the lower polariton dispersion. The black circles on the line are the signal pump and idler state respectively. The solid line arrows are sketching the parametric polariton scattering process. The dotted arrows show the phonon induced transfer between the idler state, the pump state and the excitonic reservoir.

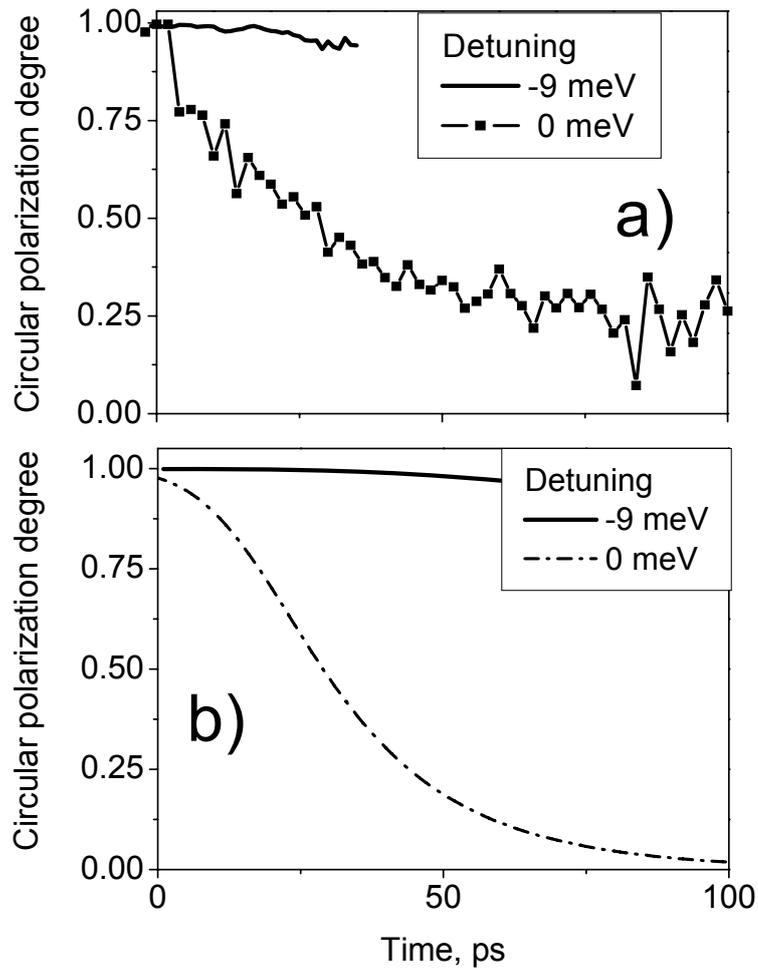

Figure 2. Circular polarization degree of the luminescence for different detuning (circular pumping, below threshold): a) experimental and b) theoretical results for sample 1 with $\tau C \approx 8$ ps .

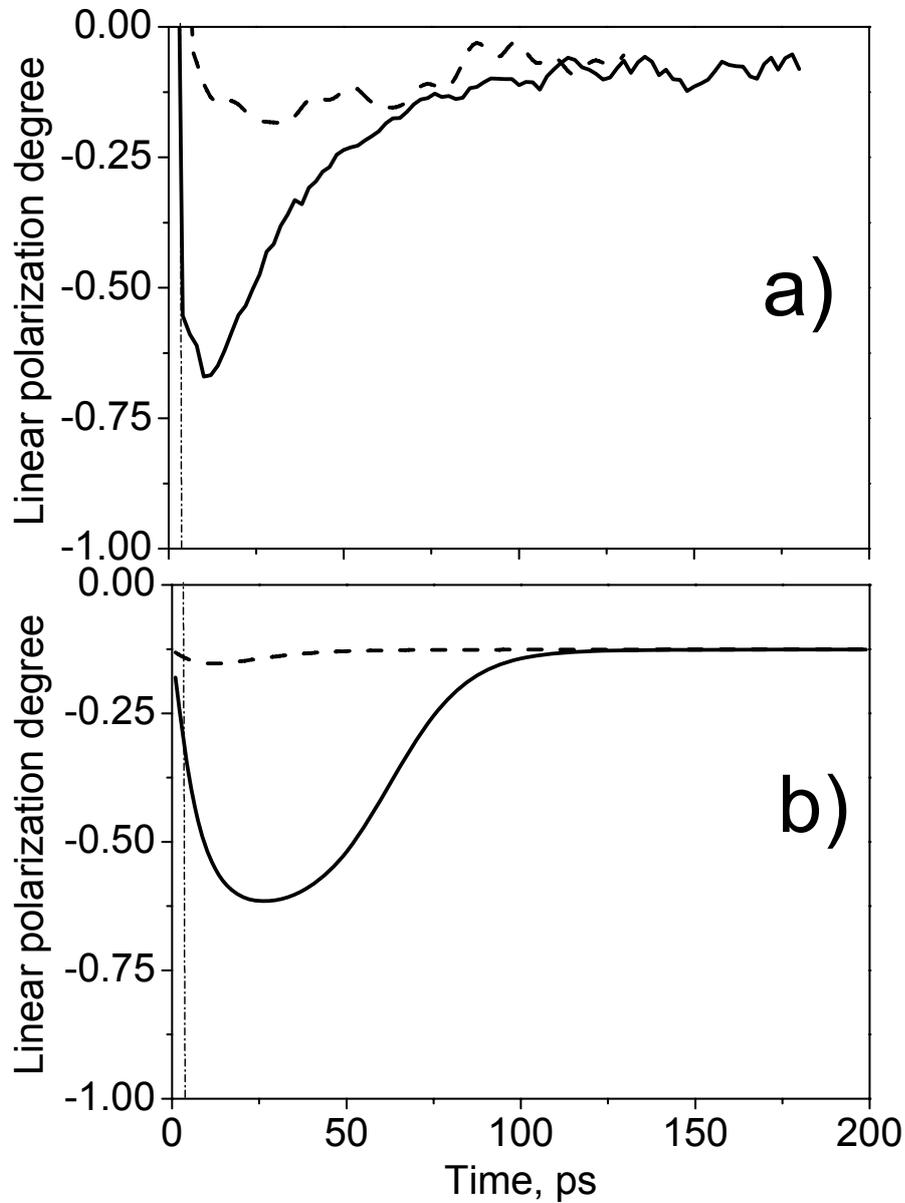

Figure 3. Linear polarization degree of the luminescence of the sample 1 for different pumping intensities (linear pumping): a) experiment, b) theory. The dashed line corresponds to 0.25 W/cm$^2$ (below threshold) and solid line corresponds to 2 W/cm$^2$ (above threshold). (Note that the initial polarization positive peak in (a) is due to diffusion of the excitation pulse on the sample surface).

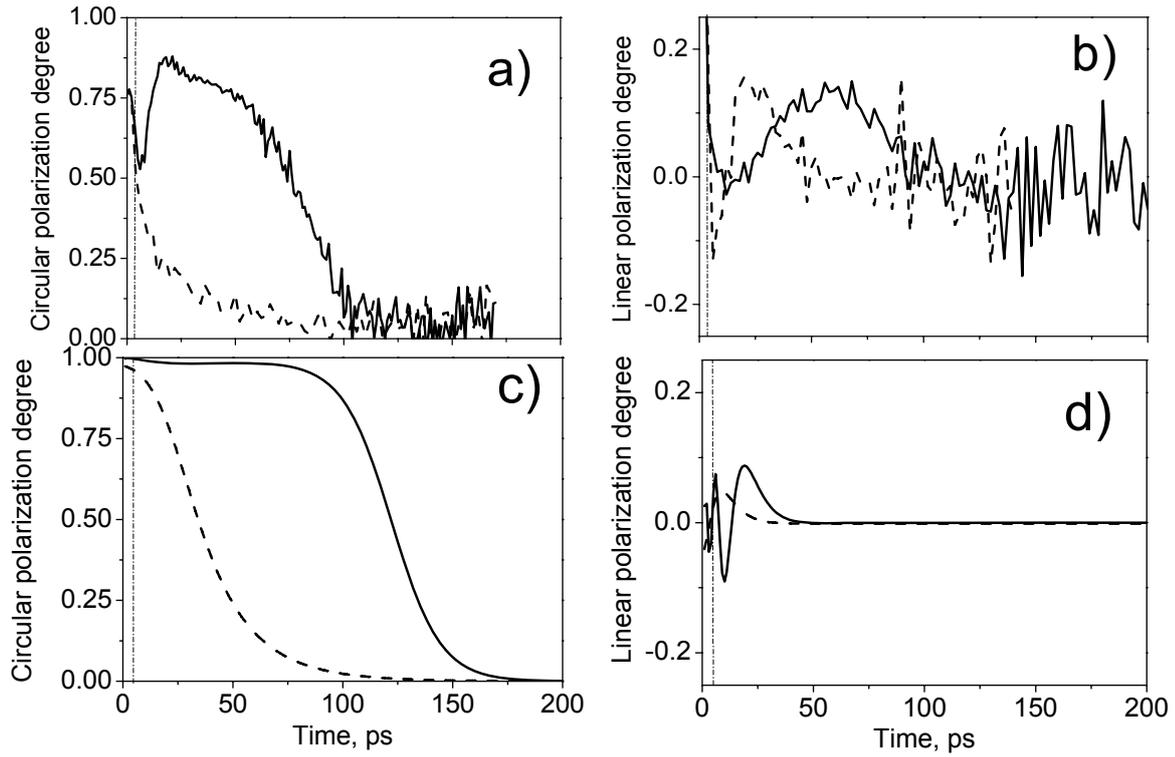

Figure 4. Circular and linear polarization degree of the signal emission for elliptically polarized pumping: experiment (a, b) and theory (c, d). Dashed line corresponds to 0.5 W/cm$^2$ (below threshold) and solid line corresponds to 3 W/cm$^2$ (above threshold). Note that the initial polarization peak in (a,b) is due to diffusion of the excitation pulse on the sample surface.